\documentclass[final]{aipproc}
\usepackage{epsfig}

\layoutstyle{6x9}

\def\kpipi{$D^+\rightarrow K^-\pi^+\pi^+$\ }

\def\ka14{$K^*_0(1430)$\ }
\def\d3pi{$D^+\rightarrow \pi^-\pi^+\pi^+$\ }
\def\ds3pi{$D_s^+\rightarrow \pi^-\pi^+\pi^+$\ }
\def\f2{$f_2(1270)$\ }
\begin{document}

\title{Phase Motion of the Scalar   $\pi \pi$ Amplitudes \\
 in $D^+$, \ds3pi Decays \footnote{Talk presented at the Scalar Meson Workshop,
 May 2003, SUNYIT, Utica, NY}}

\author{Ignacio Bediaga}{
  address={ Centro Brasileiro de Pesquisas F\'\i sicas-CBPF\\
  Rua Xavier Sigaud 150, 22290-180 Rio de Janeiro, Brazil\\
  ~ \\
  Representing the Fermilab E791 collaboration.}
}

\begin{abstract}
  We make a direct and model-independent measurement of the 
low $\pi^+\pi^-$ mass   phase  motion in the $D^+ \to \pi^-\pi^+\pi^+$ decay.
 Our preliminary results show  a strong phase variation,    compatible with 
 the isoscalar  $\sigma(500)$ meson.  This result confirms  our previous result \cite{prl} where we 
found evidence for the existence of  this scalar particle using full Dalitz-plot
 analysis. We apply   the  Amplitude Difference (AD) method \cite{ad} to 
the same Fermilab  E791  data sample used  in the preceding  analysis. 
We also give an example of how we extract the phase motion of the scalar 
amplitude, looking at  the $f_0(980)$ in \ds3pi decay. 
\end{abstract}

\maketitle


\section{Introduction}

 Fermilab experiment E791, with a full Dalitz plot analysis, showed 
strong evidence for the existence of  light and broad scalar 
resonances in charm $D^+$ meson decay \cite{prl,kappa}. 
The $\pi^+ \pi^-$ resonance is compatible with the  isoscalar meson $\sigma(500)$, 
and was observed  in the Cabbibo-suppressed decay \d3pi.  To get a good 
 fit quality  in this analysis, it was necessary to  include an
  extra scalar particle, other than the well established dipion 
  resonances \cite{pdg}. For the new state,  modeled by a Breit-Wigner 
  amplitude,  it was measured a mass and width of  $478^{+24}_{-23} \pm 17 $ MeV/$c^2$ 
  and  $324^{+42}_{-40} \pm 21 $ MeV/$c^2$respectively . The 
  $D^+\to\sigma(500)\pi^+$ decay 
  contribution is dominant, accounting for  approximately half of this 
  particular \d3pi  decay.  We found also  evidence for a  scalar 
  $ K^-\pi^+$ resonance, or  $\kappa$,  in the 
  Cabibbo-allowed decay  \kpipi \cite{kappa}. Further studies about $\kappa$  are discussed 
  in this proceeding  \cite{gobel}.

In full Dalitz plot  analyses, each possible resonance  amplitude is  
 represented by a Breit-Wigner function multiplied by angular distributions
 associated with the spin of the resonance. 
 The various contributions are combined in a coherent sum with complex
 coefficients that are  extracted from maximum likelihood fits to the data.
  The absolute value of the coefficients are related to the relative 
  fraction of each contribution and the phases take into account the final 
  state interaction (FSI) between the resonance and the third pion.

 Due to the importance of this scalar meson in many 
areas of particle and nuclear physics, it is desirable 
to be able to confirm the amplitude's phase motion in   a direct observation, 
without having to assume, a priori, the Breit-Wigner 
phase approximation for low-mass and broad resonances
 \cite{ochs, oller, polosa}. Recently, a method was proposed to 
extract  the phase motion of a complex amplitude in three 
body heavy meson decays \cite{ad}. The phase variation of a 
complex amplitude can be directly revealed through the interference 
 in the Dalitz-plot region where it crosses with a well established 
 resonant state, represented by a Breit-Wigner.

Here we begin with a simple example, showing  that the AD method  
can be applied to  extract  the  resonant phase motion of   the scalar 
amplitude  due to the  resonance $f_0(980)$,   using the same $f_0(980)$   resonance  in the 
crossing channel in  the Dalitz plot of the decay \ds3pi using 
 E791 data \cite{prlds}. This example 
shows  the ability of   this method to extract the phase motion  of an
amplitude. Then  we apply the AD method  
 using the  well known  $f_2(1270)$ tensor meson in the crossing channel, 
 as the base resonance, to   extract  the phase motion of the scalar low mass
 $\pi\pi$ amplitude in \d3pi, confirming the $\sigma(500)$ suggested by the 
 E791 full Dalitz plot analysis \cite{prl}.

\section{Extracting  $\lowercase{f}_0(980)$ phase motion with the AD method.}

\begin{figure}[t]
\epsfxsize=20pc
\centerline{
\epsfbox{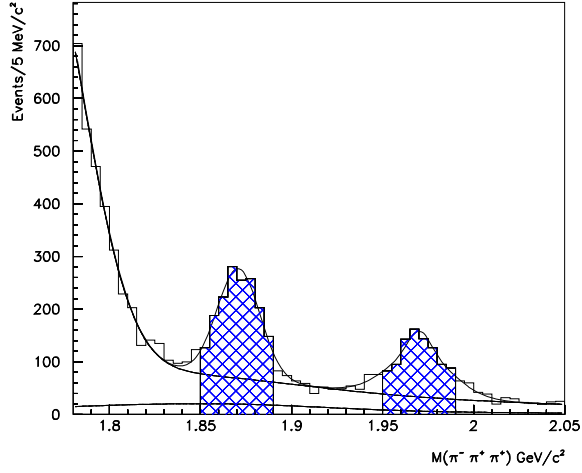}} 
\caption{The $\pi^-\pi^+\pi^+$ invariant mass spectrum. The  dashed line 
represents the total background. Events used for the Dalitz analyses
are in the hatched areas.}
\label{m3pi} 
\end{figure}

\begin{figure}[t]
\centerline{
\begin{minipage}{3.5in}
\epsfxsize=19pc
\epsfbox{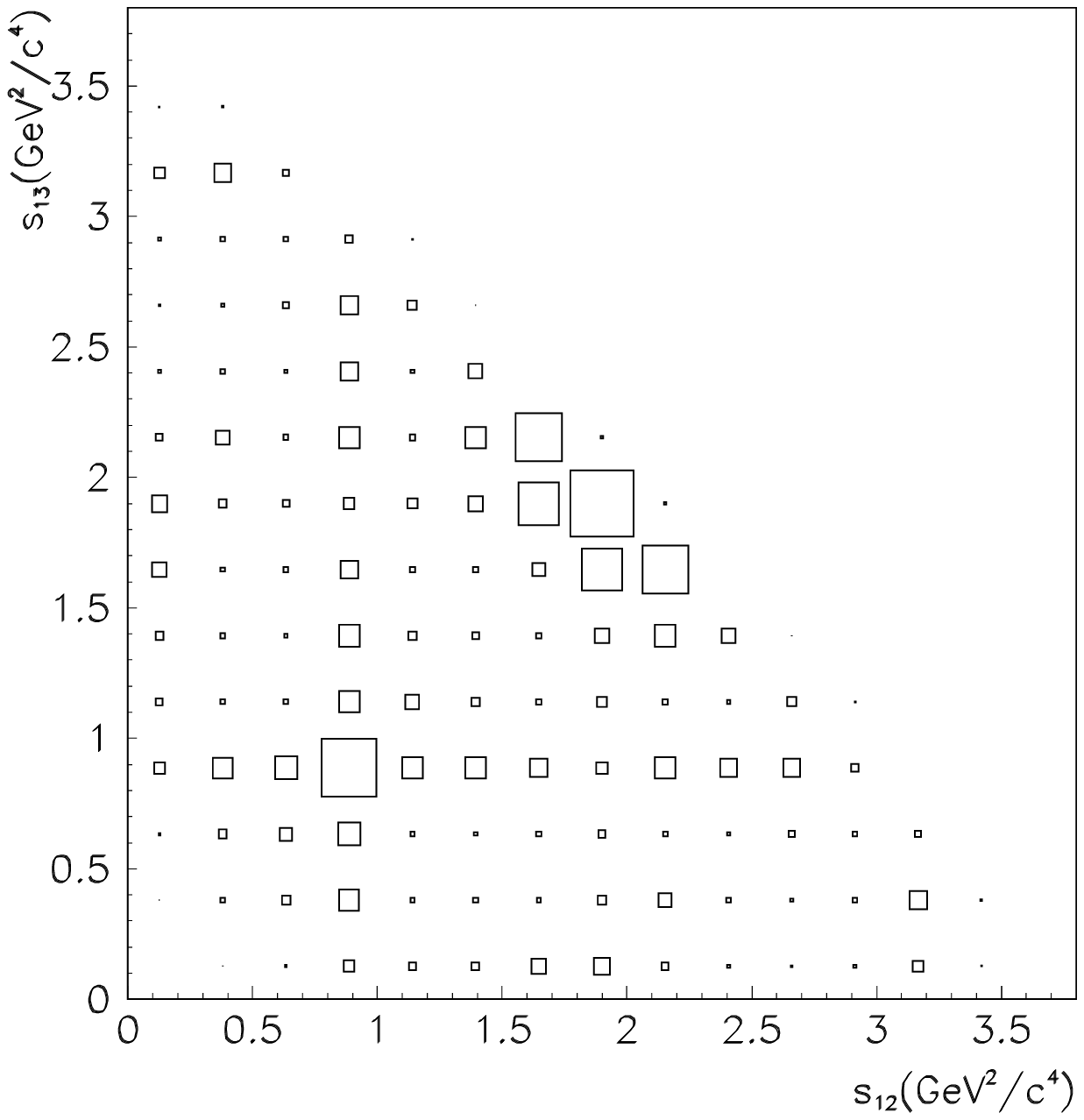}
\end{minipage}
\begin{minipage}{3.5in}
\epsfxsize=20pc
\epsfbox{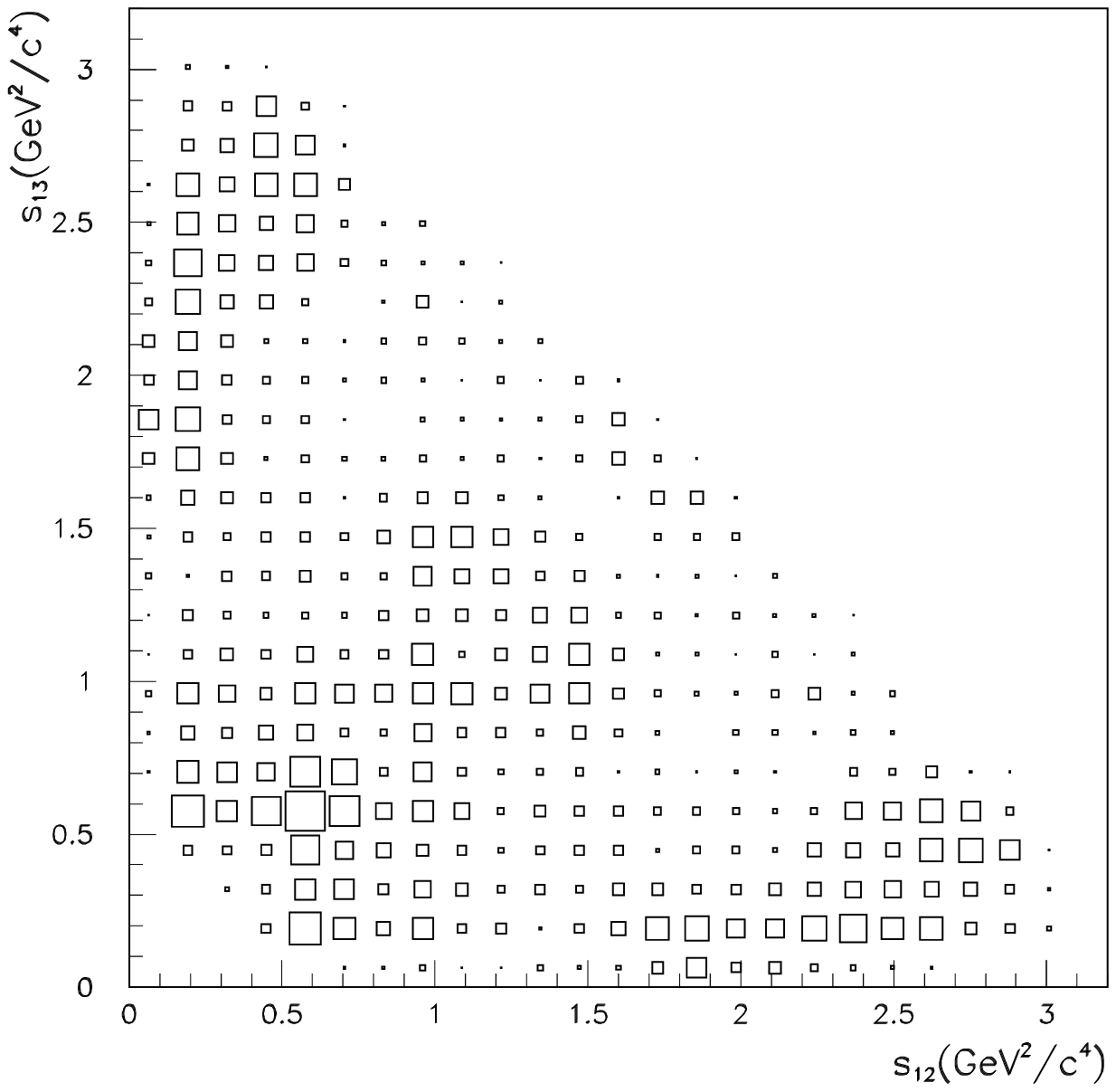}
\end{minipage}}
\caption{(a) The $D_s^+ \to \pi^- \pi^+ \pi^+$ Dalitz plot and
(b) the \d3pi Dalitz plot. Since there are two identical pions, 
the plots are symmetrical.
\label{dalitz3pi} } 
\end{figure}

From the original $2\times 10^{10}$ event  data 
collected in 1991/92 by Fermilab experiment E791 from $500
  GeV/c$  $\pi^--nucleon$ interactions \cite{ref791}, and after 
reconstruction and selection criteria, we obtained the $\pi^-\pi^+\pi^+$
 sample shown in Figure~\ref{m3pi}. To study the resonant structure 
of these three-body  decays  we consider the 1686  events with invariant mass between 
1.85 and 1.89 GeV/c$^2$, for the $D^+ $ analysis and the 937 events with invariant 
mass between 1.95 and 1.99 GeV/c$^2$ for the $D^+_s $.  
Figure~\ref{dalitz3pi}(a) shows the Dalitz-plot for the \ds3pi selected events and 
Figure~\ref{dalitz3pi}(b) the Dalitz-plot for \d3pi events. 
The two axes are the squared invariant-mass combinations for $\pi^-\pi^+$, 
and the plot is symmetrical with respect to the two identical pions.

We can see in Figure~\ref{dalitz3pi}(a)  the scalar $f_0(980)$ 
in $s_{12}$, the  square invariant mass,  crossing the $f_0(980)$  in  $s_{13}$, 
forming an interference region around  $s_{13} = s_{12} =$ 0.95GeV$^2 $. 
The AD method uses the interference region, between two 
crossing resonances, to extract the phase motion of one of them, and Final
 State Interaction (FSI) phase, provided that the second is represented by a
 Breit-Wigner \cite{ad}. In fact we are using a  {\it Bootstrap} approach;
 that is, using   a well established resonance $f_0(980)$ in $s_{12}$ to 
 extract its 
 phase motion  in $s_{13}$. It is a nice and didactic 
 example to show the ability of this method to extract  the phase 
 motion of an amplitude and the FSI phase within the E791 data sample.

The coherent amplitude to describe the crossing between a well known
scalar resonance, represented by a Breit-Wigner in $s_{12}$, and a 
 complex  amplitude under study in $s_{13}$  in a limited region of the
  phase space, where we can neglect  any other  contributions, is given 
  by:  
 
\begin{equation}
 {\cal A}(s_{12},s_{13}) = a_R {\cal BW}(s_{12}) 
+ a_s/(p^*/\sqrt s_{13}) sin \delta (s_{13}) e^{i(\delta(s_{13})+\gamma)} 
\end{equation}

$p^*/\sqrt s_{13}$ is a phase space factor to make this description compatible
with $\pi \pi$ scattering, $\gamma$ is the final state 
interaction (FSI) phase difference between the two amplitudes, $a_R$ and $a_s$ are 
respectively the real magnitudes of the resonance and the under-study complex 
amplitude. Finally $sin \delta (s_{13}) e^{i\delta(s_{13})}$ represents 
 the most general amplitude for a two-body hadronic interaction. 
   
The Breit Wigner distribution is given by:  

\[  {\cal BW} = {m_0 \Gamma_0 \over {m_0^2 - s - im_0\Gamma(m)}} \]

Taking the amplitude square of Equation 1 we get:

\begin{eqnarray} 
\mid {\cal A}(s_{12},s_{13})\mid^2  = 
a_R^2 \mid {\cal BW}_{f_0(980)}(s_{12}) \mid^2 +  
a_s^2 /p^{*2}/s_{13}\hspace{.2cm} sin^2\delta(s_{13}) \nonumber \\ 
+{2 a_R a_s m_0 \Gamma_0 sin\delta(s_{13}) {/(p^*/\sqrt s_{13})}  
\over  (m_0^2-s_{12})^2 + m^2_0 \Gamma^2(s_{12})}\hspace{.1cm} \times 
\hspace{.1cm}
[(m_0^2-s_{12}) cos(\delta(s_{13}) + \gamma) + m_0 \Gamma_0 sin(\delta(s_{13}) + 
\gamma)] 
\end{eqnarray} 

Since the Breit-Wigner is approximately symmetrical  around 
$m_0$ as seen in Figure  ~\ref{f0} (the asymmetries would come from 
$\Gamma(m)$, and is negligible for the narrow $f_0(980)$). We can divide our 
$f_0(980)$ mass  distribution in two pieces, one for  $ m_0+\epsilon$ and the 
other with  $m_0-\epsilon$. 
From Equation 2 and noticing that the pure Breit-Wigner term will cancel 
we can write:

\begin{figure}[t]
\epsfxsize=17pc
\centerline{
\epsfbox{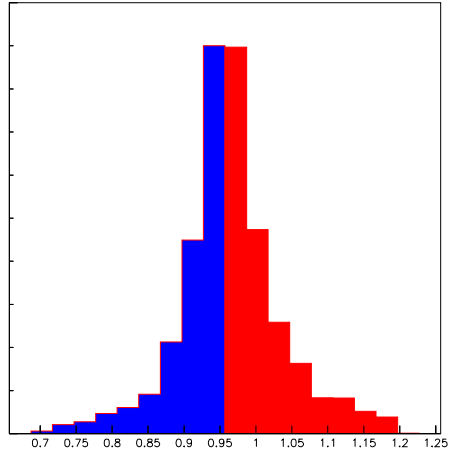}} 
\caption{$f_0(980)$ $s_{12}$  distribution, divided in 
 $m_0+\epsilon$ (red) and $m_0-\epsilon$ (blue) }
\label{f0} 
\end{figure} 

\begin{eqnarray}
\mid {\cal A}( m_0^2 + \epsilon, s_{13} ) \mid^2 
-\mid {\cal A}( m_0^2 - \epsilon, s_{13}) \mid^2  = 
{{- 4 a_s a_R/(p^*/\sqrt s_{13}) \epsilon m_0 \Gamma_0 \over\epsilon^2 + 
m_0^2\Gamma_0^2}(sin(2 \delta(s_{13})+ \gamma) - sin \gamma)} 
\end{eqnarray} 
Only the real part of the interference term in Equation 2 remains.

\begin{figure}[hbt]
\centerline{\epsfig{file=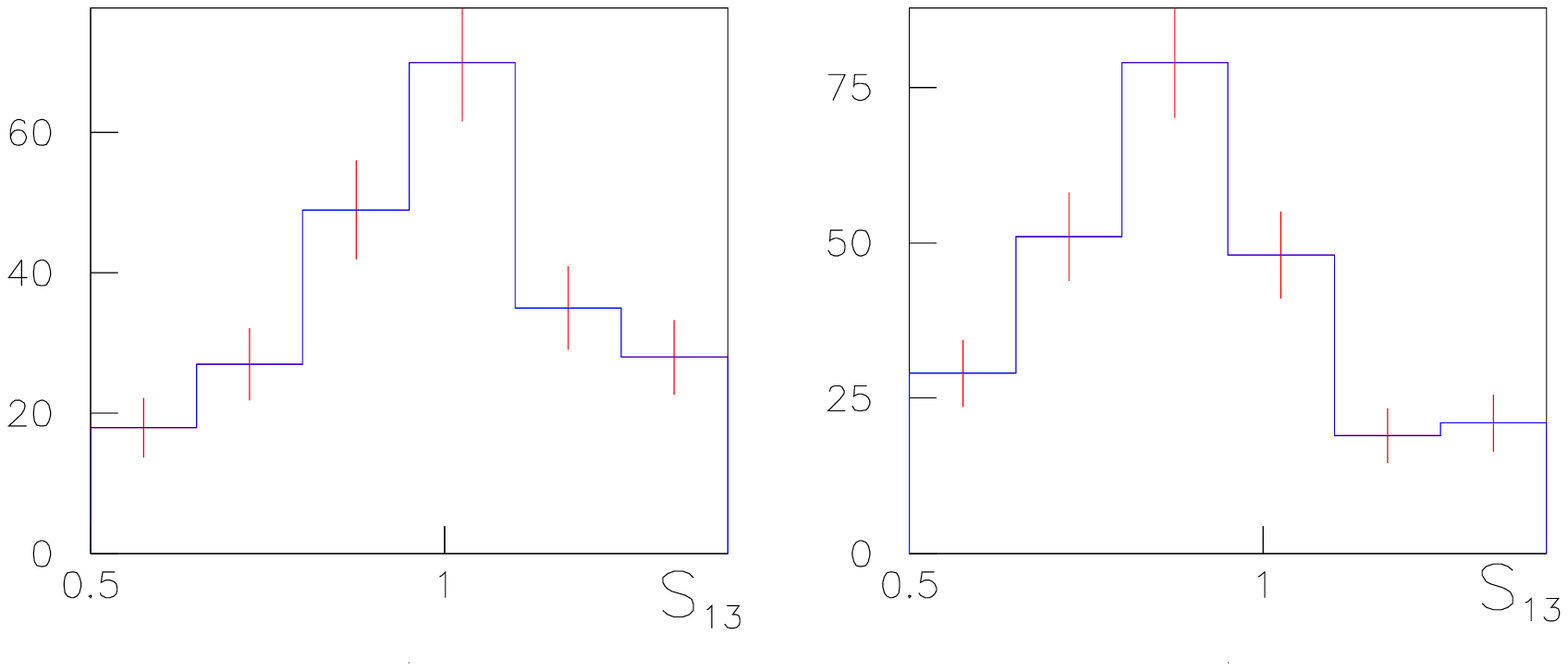,width=12 cm}}
\caption{$s_{13}$ distribution. a) For events $\int_{m^2_0}^{m^2_0+\epsilon}\mid {\cal A}(s_{12},s_{13})\mid^2 
ds_{12}$. b) For events ${\int_{m^2_0-\epsilon}^{m^2_0}\mid {\cal A}(s_{12},
s_{13})\mid^2 ds_{12}}$.}
\label{data1_f0xf0}
\end{figure}

 To extract the phase motion of the scalar amplitude in $s_{13}$ 
 through the $f_0(980)$  in $s_{12}$, represented by a Breit-Wigner,
 we took the events in $s_{12}$ between 0.7 and 1.2 GeV$^2$ and divided 
 them into two bins, as presented in Figure  ~\ref{f0}. 
 The $s_{13}$  distribution for the events of the $s_{12}$ region
integrated  between  0.95 and 1.2 GeV$^2$, is shown in 
Fig. \ref{data1_f0xf0}a and the same in   Figure  \ref{data1_f0xf0}b 
 for events  integrated between  0.7 and 0.95 GeV$^2$.
 
We can see that the peaks  in these two plots are in 
different $s_{13}$ positions. The subtraction of these distributions,
corresponds to the integration of Equation 3, that we can write as:

\begin{eqnarray}
 \Delta  \int { \cal A}^2 = 
\int_{m_{eff}^2}^{m_{eff}^2 + \epsilon}\mid {\cal A}( s_{12}, s_{13} ) \mid^2 ds_{12}
-\int_{m_{eff}^2 - \epsilon}^{m_{eff}^2}\mid {\cal A}( s_{12}, s_{13}) \mid^2 
ds_{12} \nonumber \\ 
\sim { - {\cal C}(  sin(2 \delta(s_{13})+ \gamma) - sin \gamma)}
\end{eqnarray}
where ${\cal C}$ is a constant  factor coming from the constant and integrated
 factors of Equation 3,  to be determined from data. The variation of the 
 phase space in the integral  was 
 considered  negligible for the $f_0(980)$ resonance. $ \Delta  \int { \cal A}^2 $
 directly reflects the behavior of $\delta(s_{13})$.  A constant 
 $\Delta\mid{ \cal A}\mid^2$ would imply constant $\delta(s_{13})$. This would
 be the case for a non-resonant contribution. The same way a slow phase motion
 will produce a slowly  varying $\Delta\mid{\cal A}\mid^2$ and a full resonance
 phase motion produces a clear signature in $\Delta\mid{\cal A}\mid^2$ with the
 presence of zero, maximum and minimum values.

\begin{figure}[hbt]
\centerline{\epsfig{file=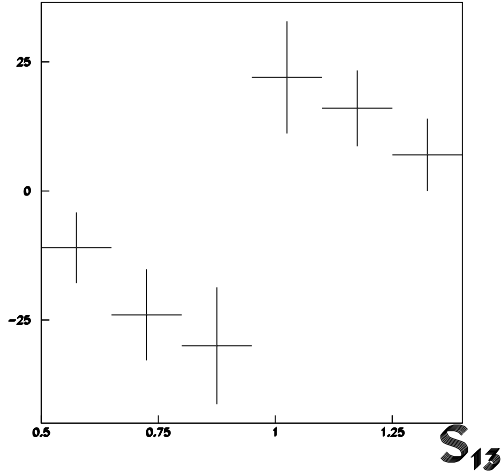,width=7 cm}}
\caption{ $s_{13}$ distribution  of $ \Delta \int  { \cal A}^2 ds_{12}$.}
\label{data2_f0xf0}
\end{figure}

The subtracted distribution, corresponding to Equation 4,  is shown 
in Figure  \ref{data2_f0xf0}. There  is a significant difference between the 
minimum (bin3) and maximum (bin4) of $ \Delta \int { \cal A}^2 $. 


We can see in Equation 4  that the zeros  occur when 
$\delta(s_{13})$ = $0^0$, $180^0$ or  $\pi/2 -\gamma$. 
In Figure  \ref{data2_f0xf0} we can see  a zero at $s_{13}$ 
near 0.5GeV$^2$, another one at $s_{13}$ = 0.95GeV$^2$ and 
a third zero near 1.4GeV$^2$. Assuming $\delta( s_{13})$ is an analytical 
function of $ s_{13}$, Equation 4 allow us to set the two following 
conditions at the maximum and minimum values of 
$\Delta\int{ \cal A}^2$ respectively:

\begin{equation}
\Delta\int{ \cal A}^2_{max} \rightarrow sin(2\delta (s_{13})+\gamma) = -1
\end{equation}

\begin{equation}
\Delta\int{ \cal A}^2_{min} \rightarrow sin(2\delta (s_{13})+ \gamma) = 1
\end{equation}

with these two conditions  we get  ${\cal C}$ and
$\gamma$, calculated from the maximum and minimum values of the
$\Delta \int {\cal A}^2$ distribution in Figure  \ref{data2_f0xf0}:

\begin{equation}
 {\cal C} = (\Delta \int {\cal A}^2_{max}-\Delta \int{\cal A}^2_{min})/2 
\end{equation}

\begin{equation}
\gamma = sin^{-1} ({\Delta \int { \cal A}^2_{max}  + \Delta\int { \cal A}^2_{min} 
\over \Delta \int { \cal A}^2_{min}  -\Delta\int  { \cal A}^2_{max} } )
\end{equation}

From  Figure  \ref{data2_f0xf0} and using the equations above, we measure  
$\gamma =   -0.15 \pm  0.31$, that is compatible with zero, as  should be 
  since we are crossing  the same resonances with, of course, the same 
final  state  interaction phase. 

With the above  conditions we solve Equation 4 for $\delta( s_{13})$:

\begin{equation}
\delta (s_{13}) = {1\over 2} ( sin^{-1} ({1 \over { \cal C}} 
\Delta\mid { \cal A}( s_{13})\mid^2 + sin(\gamma)) - \gamma )
\end{equation}

\begin{figure}[hbt]
\centerline{\epsfig{file=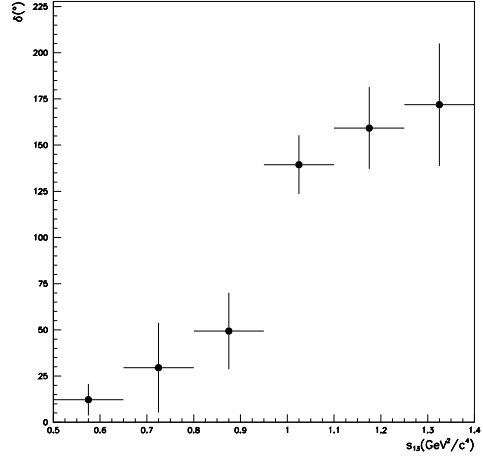,width=7 cm}}
\caption{ $\delta(s_{13})$ plot with the errors.}
\label{f0dlt}
\end{figure}

Assuming that  $\delta( s_{13})$  is an  increasing  function of  $ s_{13}$, 
 we can extract directly the $\delta( s_{13})$ value  from each bin  of 
 Figure  \ref{data2_f0xf0}, creating the $f_0(980)$ phase motion shown in Figure 
  \ref{f0dlt}. The errors in the
 plot  were produced by generating  statistically compatible experiments,  
 allowing each bin of  $\int_{m^2_0}^{m^2_0+\epsilon}\mid {\cal A}(s_{12},s_{13})\mid^2$  
(Figure\ref{data1_f0xf0}a) and 
$\int_{m^2_0-\epsilon}^{m^2_0}\mid {\cal A}(s_{12},s_{13})\mid^2$
 (Figure\ref{data1_f0xf0}b) to fluctuate randomly following a Poisson law. We 
 then solve the problem for each "experiment". The error in each bin of  $\delta(s_{13})$ 
 will be  the RMS of the distributions generated by the "experiments".

From Figure \ref{f0dlt} we can see what one could expect, that is the scalar 
amplitude near 970GeV  with a phase motion of about $180^0$ degrees. 
This example demonstrates the ability of  AD method to extract the phase motion 
 of an amplitude   with E791 statistics.

\section{Extracting the scalar low mass $\pi \pi$ amplitude  phase motion 
with the AD method.}

In the preceding  section, we showed how to apply the AD method to extract the 
phase motion of an amplitude, from nonleptonic charm-meson three-body decay.  
Here we apply the same method to extract the phase motion of the 
scalar low-mass $\pi\pi$ amplitude  in \d3pi decay, where we previously found  
 strong experimental evidence for the existence of  a light and broad 
isoscalar resonance \cite{prl}. To start 
this analysis,  we have to decide what is the best well-known resonance to be used 
for crossing the low mass amplitude under study. Taking a look at 
Figure~\ref{dalitz3pi}b 
we can see the signature of three resonances that in principle could be used, 
the  $\rho(770)$, $f_0(980)$ and $f_2(1270)$. In fact, the E791 analysis of this 
Dalitz plot found  
a significant contribution from these three resonances in \d3pi decay  \cite{prl}.  
Since this $D^+$  decay is symmetric for the exchange of the $\pi^+$ meson, 
each resonance 
in $s_{12}$ is present also in  $s_{13}$. Then if we use  $\rho(770)$ as the 
base resonance in $s_{12}$, we have  also the presence of the $\rho(770)$ in same 
mass square distribution of the $\sigma(500)$ in $s_{13}$. The  proximity of the  
$\rho(770)$ with the $\sigma(500)$, both  broad resonances, creates an overlap 
between them such that we are not able to separate the phase motion of one from the other. We could use the $f_0(980)$ 
as a base resonance, but again the presence of the $\rho(770)$ overlapping with the 
$\sigma(500)$ creates the same problem. 

\begin{figure}[hbt]
\centerline{\epsfig{file=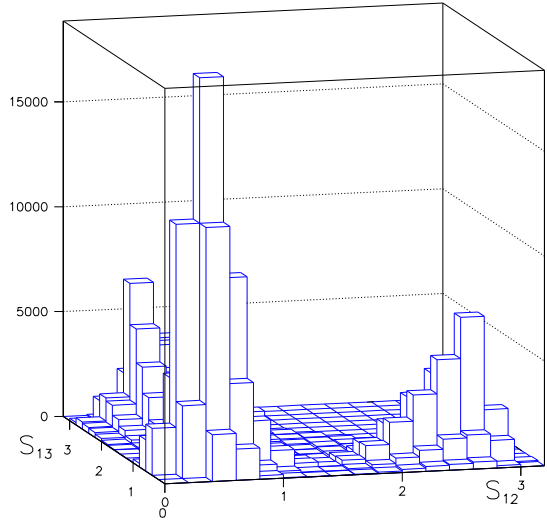,width=8 cm}}
\caption{ MC $\rho(770)$ distribution in \d3pi decay.
There is  little contribution between 1.2 to 1.8GeV$^2$ }
\label{rho}
\end{figure}

There remains only the tensor meson $f_2(1270)$ candidate at 
$m_0^2 = 1.61GeV^2$,  which is placed   where the  $\rho(770)$ contribution 
 reaches  a minimum due the angular distribution  in the  middle of the 
 Dalitz plot,  as we can see from  the $D^+\to \rho(770) \pi^+$ decay 
 Monte Carlo simulation shown  in  Figure~\ref{rho}.

  The   amplitude for the crossing of the $f_2(1270)$ in  $s_{12}$ and the   
  complex amplitude  under study in $s_{13}$ is given in the same way
   as in Equation 1:

\begin{eqnarray}
{\cal A}(s_{12},s_{13}) = a_{R}\hspace{.3cm}  {\cal
BW}_{f_2(1270)}(s_{12}) {
\hspace{.2cm}   ^{j=2}{\cal M}_{f_2(1270)}(s_{12},s_{13})} + \\ 
+\hspace{.3cm} a_s/(p^*/\sqrt s_{13}) \hspace{.2cm} sin \delta (s_{13}) \hspace{.3cm} 
e^{i(\delta(s_{13})+\gamma)}
\nonumber
\end{eqnarray}
where $^{j=2}{\cal M}_{f_2(1270)}(s_{12},s_{13})$ is the angular 
function for the $f_2(1270)$ tensor resonance. The amplitude under study 
represents the scalar low mass $\pi\pi$ amplitude  in a limited region 
of the phase space, where we can neglect  the  other amplitude contributions.

\begin{figure}[hbt]
\centerline{\epsfig{file=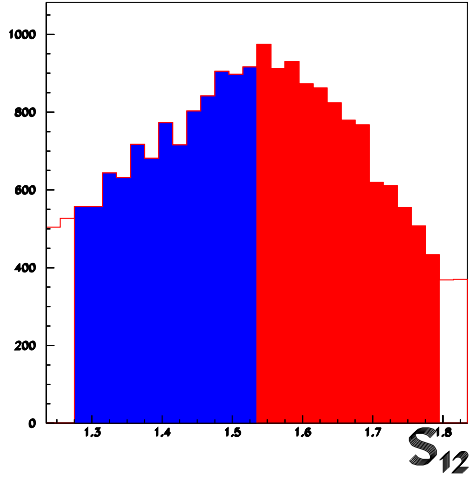,width=7 cm}}
\caption{ Monte Carlo $f_2(1270)$ $s_{12}$  distribution, divided in 
 $m_0+\epsilon$ (red) and $m_0-\epsilon$ (blue)}
\label{f2_s12}
\end{figure}

Both the   width  $\Gamma(s_{12})$  and  the angular function
  $^{j=2}{\cal M}_{f_2(1270)}$ from this resonance  produce  asymmetries in 
 $s_{12}$ and  consequently  we can  not use the nominal $f_2(1270)$   mass to 
 divide our sample into two slices, as we did for the $f_0(980)$ example. So we performed a
 Monte Carlo study to determine the effective mass we must  use. The $s_{12}$ 
  Monte Carlo projection of the $f_2(1270)$ in \d3pi decay is 
shown  in Fig. \ref{f2_s12}. We can see the asymmetry created around 
the nominal $f_2(1270)$   mass value due to $\Gamma(s_{12})$  and  
 $^{j=2}{\cal M}_{f_2(1270)}$ contributions to the amplitude.  

Here  we require   an effective mass squared ($m_{eff}$), such that the number 
of events integrated between  $ m_{eff}^2$ and $ m_{eff}^2 + \epsilon$  is 
equal, by construction, to the number of events integrated between  
$ m_{eff}^2$ and $ m_{eff}^2 - \epsilon$. 
We choose, using the $f_2(1270)$   Monte Carlo distribution,  a mass  
of { $m_{eff}^2 = 1.535GeV^2$}, within   { $  \pm 0.26 GeV^2 $} 
\footnote {Within this mass region,  the amount of $\rho(770)$ events 
was estimate to be  around 5\%}, in such  way that  we can write: 

\begin{equation}
\int_{m_{eff}^2}^{m_{eff}^2 + \epsilon}
 \mid{\cal BW}_{f_2(1270)}(s_{12})  
 { ^{j=2}{\cal M}_{f_2(1270)}}\mid^2 ds_{12}=
 \int_{m_{eff}^2 - \epsilon}^{m_{eff}^2}
 \mid{\cal BW}_{f_2(1270)}(s_{12}) 
 { ^{j=2}{\cal M}_{f_2(1270)}}  \mid^2 ds_{12}
\end{equation}

The effective mass squared $m_{eff}$ and  the separation between  
$ m_{eff}^2 + \epsilon$ (red) and $ m_{eff}^2 - \epsilon$ (blue) are 
shown in   Figure \ref{f2_s12}.

\begin{figure}[hbt]
\centerline{\epsfig{file=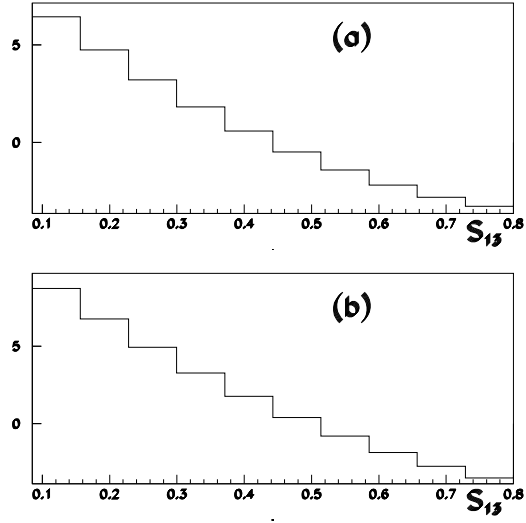,width=8 cm}}
\caption{ Fast MC $^{j=2}{\cal M}_{f_2(1270)}$ distribution in $s_{13}$. a) For events 
between  $ m_{eff}^2$ and $ m_{eff}^2 + \epsilon$. b) For events 
between $ m_{eff}^2$ and $ m_{eff}^2 - \epsilon$.}
\label{f2_s13}
\end{figure}

 The    $^{j=2}{\cal M}_{f_2(1270)}$ function in  $s_{13}$  
  is presented  in Figure \ref{f2_s13}\footnote{ Since 
  we divided our data sample by this function, we represent this function 
  in a histogram with the same binning of data.}. The distribution between 
 $ m_{eff}^2$ and $ m_{eff}^2 + \epsilon$  is shown in  Fig. \ref{f2_s13}a,
  for events between $ m_{eff}^2$ and $ m_{eff}^2 - \epsilon$ in Fig.
  \ref{f2_s13}b.  We can see that  these two plots are slightly different. 
  However we considered  the approximation   
$ ^{j=2}{\cal M}^+_{f_2(1270)}(s_{13})\sim  \hspace{.1cm} ^{j=2}{\cal M}^-_{f_2(1270)}
(s_{13})$  and take an average function   $ ^{j=2}\bar{\cal M}_{f_2(1270)}(s_{13})$. 
Another important effect, that we had to take into account, is the zero of this function 
at $s_{13} = 0.48GeV^2$. Below we discuss the consequences of that in the AD method.

With the above considerations about the $f_2(1270)$  in $s_{12}$
and $s_{13}$ we can write the integrated amplitude-square difference as:  

\begin{eqnarray}
 \Delta  \int  {\cal A}^2  = 
\int_{m_{eff}^2}^{m_{eff}^2 + \epsilon}\mid {\cal A}( s_{12}, s_{13} ) \mid^2 ds_{12}
-\int_{m_{eff}^2 - \epsilon}^{m_{eff}^2}\mid {\cal A}( s_{12}, s_{13}) \mid^2 
ds_{12} \nonumber\\ 
\sim { - {\cal C}(  sin(2 \delta(s_{13})+ \gamma) - sin \gamma){
\hspace{.2cm}   ^{j=2}\bar{\cal M}_{f_2(1270)}(s_{13})}/(p^*/\sqrt s_{13})}
\end{eqnarray}

This Equation is similar to Equation 4, with an extra angular function term
$^{j=2}\bar{\cal M}_{f_2(1270)}(s_{13})$ \footnote{ For short we shall use, 
 from here on $^{j=2}\bar{\cal M}_{f_2(1270)}(s_{13})= \bar{\cal M} $ and
$p^*/\sqrt s_{13} = p'$.}.

 The  background and the acceptance  are similar between 
 $ m_{eff}^2$ and $ m_{eff}^2 + \epsilon$ and  
$ m_{eff}^2$ and $ m_{eff}^2 - \epsilon$. Since we are subtracting these two 
distributions, we do not take into account these  effects in our analysis.   

\begin{figure}[hbt]
\centerline{\epsfig{file=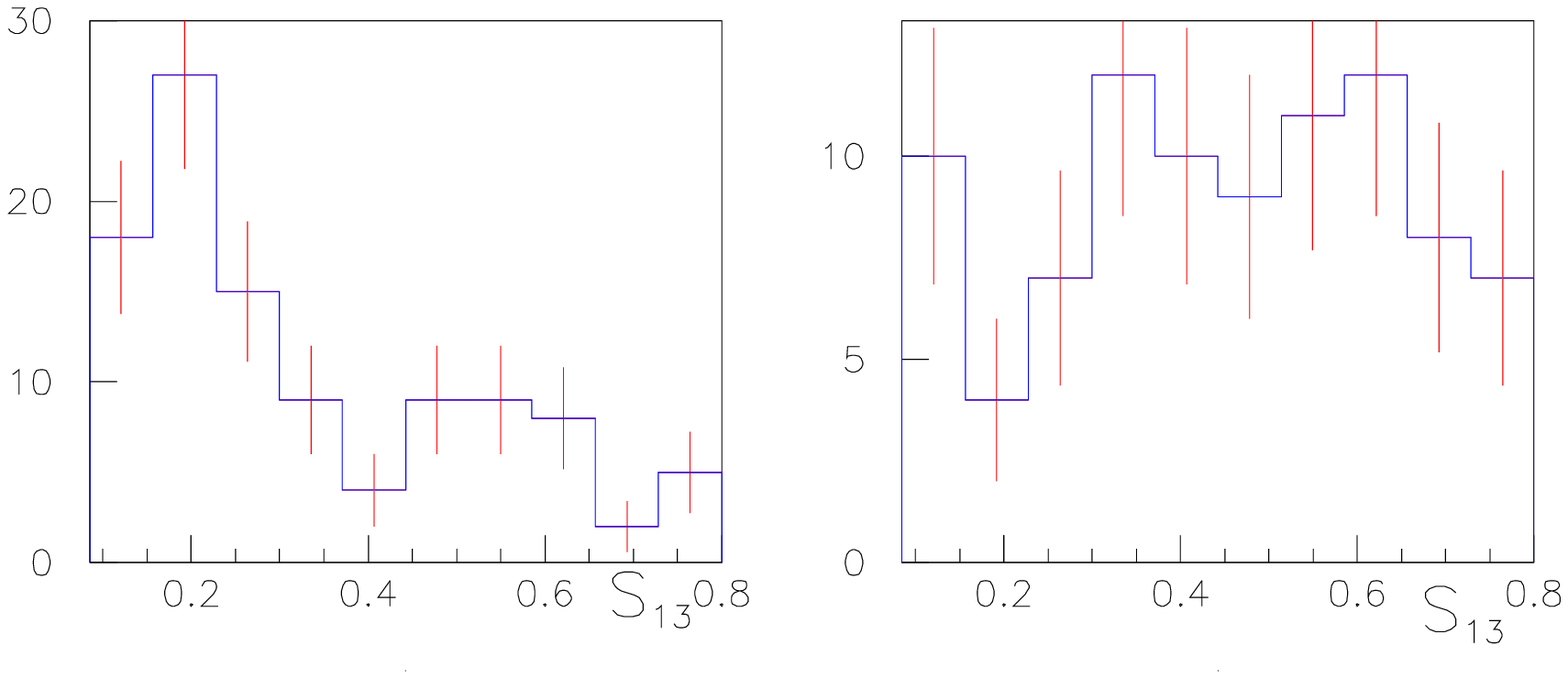,width=13 cm}}
\caption{ Events  distributions in $s_{13}$, a) for events $\int_{m^2_0}^{m^2_0+\epsilon}\mid {\cal A}(s_{12},s_{13})\mid^2 
ds_{12}$, b) For events ${\int_{m^2_0-\epsilon}^{m^2_0}\mid {\cal A}(s_{12},s_{13})\mid^2 ds_{12}}$.}
\label{data1}
\end{figure}

The  $ \int  { \cal A}^2 $  in $s_{13}$, 
for events  integrated in $s_{12}$ between $m_{eff}^2 = 1.535 GeV^2$ and  
$m_{eff}^2  +  \epsilon $ and $m_{eff}^2$  and $ m_{eff}^2  -  \epsilon $, with 
$\epsilon = 0.26 GeV^2$ are presented  in Figure 
\ref{data1}a and b respectively.

\begin{figure}[hbt]
\centerline{\epsfig{file=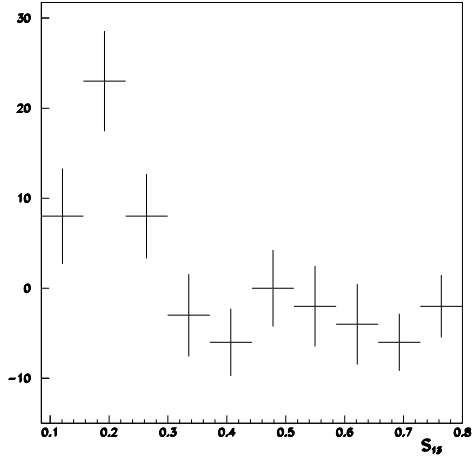,width=7 cm}}
\caption{  $s_{13}$ distribution  for $ \Delta \int  { \cal A}^2 $.}
\label{data2}
\end{figure}

Subtracting  these two histograms, in the same way we did for the 
$f_0(980)$ example, gives the $ \Delta \int  { \cal A}^2 $ of the Equation 12. 
The result is shown in Figure \ref{data2}.

\begin{figure}[hbt]
\centerline{\epsfig{file=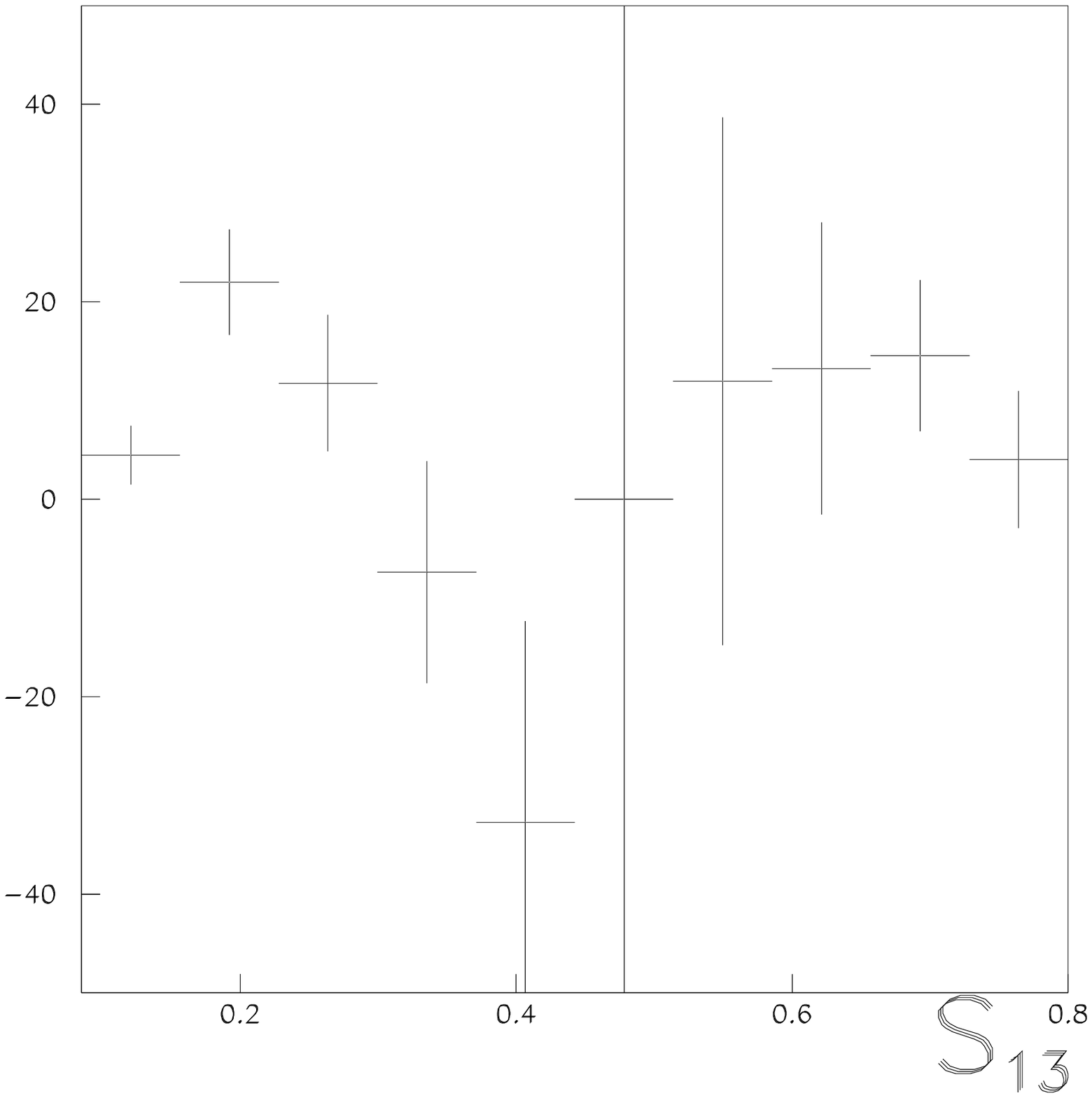,width=7 cm}}
\caption{  $s_{13}$ distribution  for 
$ \Delta \int  { \cal A}^2 \hspace{.1cm} p'/ \bar{\cal M} $.}
\label{data3}
\end{figure}

Here we can not extract directly the phase motion from Figure \ref{data2}, as 
we  did for the $f_0(980)$ example using the conditions 5 and 6. 
We have to divide the 
$ \Delta \int  { \cal A}^2 $  by   $\bar{\cal M} $ 
(average of  the distributions in Figures \ref{f2_s13}a and b) and multiply 
by $p'$, since 
phase space here is an important effect. By  doing this the only 
$s_{13}$ dependence of the right hand side of  Equation 12 is in
$\delta(s_{13})$. However, as we could see in Figure 
\ref{f2_s13}, there is a zero about 0.48GeV$^2$ in the angular function, which 
means a singularity around this value in $ \Delta \int  { \cal A}^2 /\bar{\cal M} $.
 To avoid this singularity, we first produced a binning 
in such a way that the singularity is placed in the middle of one bin. In Figure \ref{data3}
we show the  $ \Delta \int  { \cal A}^2 $  by   $\bar{\cal M} $       distribution. We can see that the 6th bin 
(around 0.48GeV$^2$), has a huge error, that corresponds to the bin 
 of the singularity. Due to the singularity we decided  not to use 
this region (bin 6) further in this analysis. The consequences of this choice
are going to be taken care of in systematic error studies.  In any case, the singularity 
could only affect the position of the minimum of Figure \ref{data3}. It does
 not interfere with the general feature of starting at zero, having 
 statistically  significant  maximum and minimum values, and coming
  back to zero,  indicating  a strong phase variation.   Bins 2 and 5  
  are respectively  the maximum and minimum value of     
   $ \Delta \int  { \cal A}^2  p'/ \bar{\cal M} $ of 
 Figure \ref{data3} where we use the Equation 5 and 6 conditions.

With the same assumptions  used for the $f_0(980)$, that is
$\delta( s_{13})$ is an analytical and increasing  function of  $ s_{13}$,
and using Equation 7, 8 and 9 (multiplied by  $p'$ and divided by $\bar{\cal M} $), 
we can extract $\gamma$ and $\delta( s_{13})$   from Figure \ref{data3}. 
For the FSI phase we found $\gamma = 3.26 \pm  0.33$, that is somewhat  
bigger than found by the E791 full Dalitz-plot analysis \cite{prl} 
($\gamma_{Dalitz} =  2.59 \pm 0.19 $).  The fact that 
we used  the effective mass for the $f_2(1270)$ = 1.535 GeV$^2$ instead of 
the nominal mass is  responsable for the shift observed in the 
relative phase. To verify this statement we generated 1000 samples of fast-MC, 
with only two amplitudes,  $f_2(1270)$ and  $\sigma(500)$. For both we 
used  Briet-Wigner functions and the E791 parameters.  We generated 
 the  phase difference of 2.59 rad, measured by the E791. For these 
1000 samples, we measure  $\gamma$ using the method presented here. 
The result has  a mean value of 3.07.  We can say that the difference 
between the generated and measured  $\gamma$ value is a correction factor 
due to the use of an effective mass. Using this offset  factor 
( 2.59 - 3.07 = -0.48) we correct the measurement  
$\gamma = 3.26 \pm 0.33 $ to  $\gamma_{corr} =  2.78 \pm 0.33 $.
So the  observed $\gamma$ difference between Dalitz analysis 
and the $\gamma_{corr}$  is in  good agreement, with a difference 
 below one standard deviation.

\begin{figure}[hbt]
\centerline{\epsfig{file=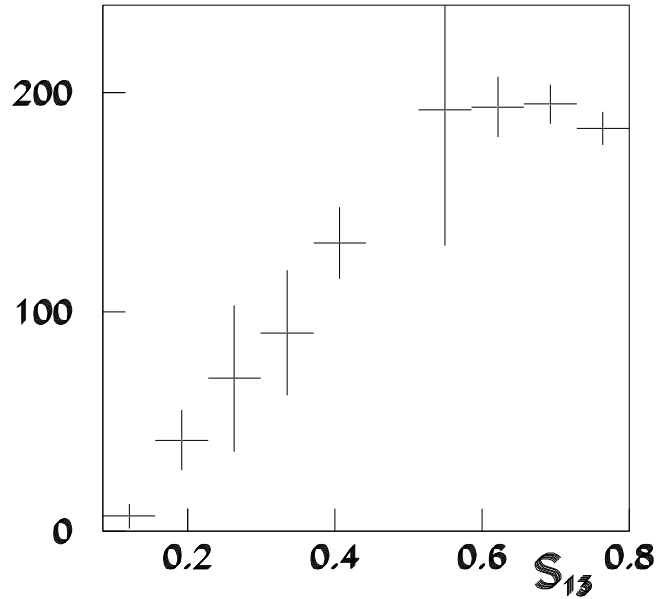,width=11 cm}}
\caption{ Phase motion $\delta(s_{13})$ distribution for the scalar low mass
 $\pi\pi$ amplitude,  with the errors.}
\label{dlt}
\end{figure}

 The  $\delta( s_{13})$ was extracted 
bin by bin, with the same approach for the errors used in the $f_0(980)$ example, and 
we got the distribution  shown in Figure \ref{dlt} \footnote{ In Figure 
\ref{dlt} there is no the 6th bin because of the singularity we mentioned  
above.}. 
We can see a strong phase variation of about 180$^0$ around the 
mass for the $\sigma(500)$, showing a phase motion compatible with a resonance. 

\section{Conclusions}

We  showed  that the AD method  can be applied to E791 data to extract  the  
phase motion of   the  resonance $f_0(980)$  in the Dalitz plot of the decay 
\ds3pi. 
This example demonstrates  the ability of   this method to extract the phase motion  
of a resonance  amplitude. 

Preliminary E791 results present  a  direct and model-independent approach,
 obtained with the AD method,  and confirms our previous result of the evidence 
 of an important contribution of the isoscalar $\sigma(500)$ meson in \d3pi decay 
 \cite{prl}. We use the well known  $f_2(1270)$ tensor meson in the crossing 
 channel, as the base resonance, to   extract  the phase motion of the 
 low mass $\pi\pi$ scalar amplitude. 
 We obtain a $\delta( s_{13})$ variation of about 180$^0$ consistent with 
  a resonant $\sigma(500)$ contribution. We also obtain  
 good agreement between the FSI $\gamma_{corr}$ observed with AD method and
 the $\gamma$ observed in the full Dalitz plot analysis.  
 
\begin{theacknowledgments}
We wish to thank the Scalar Meson Workshop  organizers, 
specially A. Fariborz, for the invitation and to offer us a pleasent 
workshop in Utica.  We would like to thank also  Wolfgang Ochs 
 for suggestions and important comments.

\end{theacknowledgments}

\end{document}